\documentclass[useAMS]{mn2e}
\usepackage{times}
\input{epsf}

\title[Black hole spin in GRS 1915+105]
{Black hole spin in GRS 1915+105}

\author[M.Middleton, C. Done, M. Gierli{\'n}ski, S.W. Davis]
{Matthew Middleton$^1$, Chris Done$^1$, Marek Gierli\'nski$^{1,2}$
and Shane W. Davis$^3$\\
$^1$Department of Physics, University of Durham, South Road, Durham
DH1 3LE,
UK\\
$^2$Obserwatorium Astronomiczne Uniwersytetu Jagiello{\'n}skiego,
30-244
Krak{\'o}w, Orla 171, Poland\\
$^3$University of California at Santa Barbara, California, USA }

\pagerange{\pageref{firstpage}--\pageref{lastpage}} \pubyear{2005}

\begin{document}

\topmargin = -0.5cm

\maketitle

\label{firstpage}

\begin{abstract}

Microquasars are galactic black hole binary systems with radio jets
which can sometimes be spatially resolved to show superluminal
motion. The first and best known of this class of objects is
GRS~1915+105, the brightest accreting source in our Galaxy. There is
persistent speculation that strong jet emission could be linked to
black hole spin. If so, the high spin should also be evident in
accretion disc spectra. We search the {\it RXTE} archive to find
disc-dominated X-ray spectra from this object, as these are the only
ones which can give reliable spin determinations by this method.
Finding these is complicated by the rapid, unique limit cycle
variability, but we are able to identify such spectra by going to
the shortest possible time resolution (16 s). We fit them with a
simple multicolour disc blackbody ({\sc diskbb}), and with the best
current model which include full radiative transfer as well as
relativistic effects ({\sc bhspec}). Both these models show that the
spin is intermediate, neither zero nor maximal. {\sc bhspec}, the
most physical model, gives a value for the dimensionless spin of
$a_*\sim 0.7$ for a distance of 12.5~kpc and inclination of
$66^\circ$. This, together with the range of spins $0.1<a_*<0.8$
derived using this method for other black holes, suggests that jet
emission is probably fundamentally powered by gravity rather than
spin, and implies that high-to-maximal spin is not a pre-requisite
for powerful relativistic jets.

\end{abstract}
\begin{keywords}  X-rays: binaries -- black hole physics --
accretion, accretion discs -- stars: individual: GRS 1915+105
\end{keywords}

\section{Introduction}

Astrophysical black holes are very simple objects, completely
characterized by their mass and spin.  Mass can be measured using
Newtonian mechanics by studying objects orbiting the black hole. For
the Galactic black holes this is the companion star of the binary
system, while for the supermassive black holes in galaxy nuclei this
can be traced by nearby stars or gas in orbit around it. However,
unlike mass, spin does not leave a discernable mark on the external
spacetime at large distances.  It only makes a significant difference
close to the event horizon, so it is much more difficult to measure.

Since the spin of a black hole is generally not well known it becomes
an obvious parameter to use to `explain' any unusual behaviour. Jet
production is a case in point. About 10 per cent of quasars are known
to be radio loud, while the rest are radio quiet. Similarly, some of
the Galactic black hole binaries are also known to produce relativistic
jets (e.g. Fender, Belloni \& Gallo 2004). There is persistent
speculation in the literature that jet production is linked to the spin
of the black hole (e.g. Moderski, Sikora \& Lasota 1998).
Theoretically, this could be due to direct tapping of the spin energy
of the black hole via the Blandford-Znajek effect (Blandford \& Znajek
1977). These ideas can only be directly tested by measuring the spin,
so it becomes an important property to constrain.

The most obvious consequence of black hole spin is that it drags the
last stable orbit inwards, from 6$R_g$ (where $R_g \equiv GM/c^2$, and
$M$ is the black hole mass) for a non-rotating black hole to 1.23$R_g$
for a maximal spin of $a_*$ = 0.998, where dimensionless spin parameter
$a_*$ is defined as $Jc/GM^2$, and $J$ is the black hole angular
momentum. Thus for the same mass accretion rate through a disc, a
spinning black hole has a higher disc luminosity and temperature.
Equivalently, in terms of direct observables, a spinning black hole of
the same luminosity has a higher disc temperature. However, using this
to measure black hole spin is not entirely straightforward as black
hole spectra are generally complex. The accretion disc spectrum is
always accompanied by a tail of emission to much higher energies. While
the origin of this tail is not well understood, it clearly points to
some fraction of the energy being dissipated outside of the optically
thick disc material, so that it does not thermalize. The tail can have
a large impact on the derived disc luminosity and temperature if it
carries a large fraction of the bolometric luminosity (see e.g. Kubota
\& Done 2004), so disc radii derived from these spectra are highly
model dependent. Conversely, where the tail is not energetically
important the disc spectrum can be reliably determined, and direct
fitting of such disc dominated spectra can give an estimate for the
black hole spin (e.g. Ebisawa et al. 1991; Gierli{\'n}ski,
Macio{\l}ek-Nied{\'z}wiecki \& Ebisawa 2001). However, a further
problem is that the disc spectrum is not a simple sum of blackbody
spectra from different radii.  Distortions from Compton scattering from
within the disc itself are probably important (Shakura \& Sunyaev 1973;
Shimura \& Takahara 1995; Merloni et al. 2000; Davis et al. 2005),
making this method dependant on the assumed vertical disc structure.

A variant on this method is to use a sequence of disc dominated spectra
from a given source, at different luminosities. A constant radius disc
at different mass accretion rate should have luminosity, $L\propto T^4$
where $T$ is the maximum temperature. The proportionality constant
gives a measure of the disc area, and hence radius, modulo
uncertainties in the spectral distortions introduced by Compton
scattering in the disc itself. The advantage of this method is that
observationally many sources show $L\propto T^4$, giving confidence in
a well defined, constant inner disc radius to associate with the last
stable orbit (Ebisawa et al. 1991, 1993; Kubota et al. 2001; Kubota \&
Makashima 2004; Kubota \& Done 2004; Gierli\'{n}ski \& Done 2004), and
for an approximately constant colour temperature correction for the
spectral distortions (Shafee et al. 2006; Davis, Done \& Blaes 2006).

Here we apply this method to GRS~1915+105. This is the obvious source
to use to investigate links between black hole spin and jet behaviour
as it was the first galactic source to show a superluminal radio jet
(Mirabel \& Rodr{\'{\i}}guez 1994).  Several more superluminal jet
sources have since been found (GRO J1655--40, XTE J1748--288, V4641
Sgr), while a much larger number of black holes (together termed the
microquasars) show jet-like radio morphology (e.g. Mirabel \&
Rodr{\'{\i}}guez 1999).  Mildly relativistic jets are now recognized as
a common feature of black holes at low mass accretion rates, while
strong radio flaring is probably associated with state changes (e.g.
Fender, Belloni \& Gallo 2004).

However, GRS~1915+105 is also the most spectacularly variable accreting
black hole in our Galaxy, showing episodes where it continually
switches between states in a quasi--regular way (see e.g. the review by
Fender \& Belloni 2004 and references therein). Belloni et al. (1997a)
showed that this variability could be associated with the accretion
disc spectrum switching from hot and bright, implying a small inner
disc radius, to cooler and dimmer, with a larger inferred radius. They
interpreted this as the result of a limit-cycle instability in the
inner accretion disc, such that it is continually emptying and
refilling. Whatever its origin, this variability complicates the
spectral analysis, as does the high absorbing column density towards
this source. Nonetheless, by going to the highest resolution spectra
(16 s) we are able to find disc dominated spectra, and these show an
approximate $L\propto T^4$ relation (see also Muno et al. 1999). We fit
these spectra simultaneously with the best currently available
theoretical accretion disc models (Davis et al. 2005) and find $a_*
\sim 0.7$ for a distance of 12.5~kpc. There are considerable
uncertainties on the system parameters, but this spin is consistent
with the moderate spins found from other, better constrained, GBH jet
systems (Davis et al. 2006). We show this is also consistent with the
most recent numerical MHD simulations of the accretion flow, which show
that moderate spin black holes can produce powerful jets (Hawley \&
Krolik 2006; Mckinney 2005).

\section{Spectral model}
\label{sec:model}

We use {\sc xspec} version 11.3 for spectral fitting. The main
component of our model is emission from the optically thick accretion
disc. For data selection (Section \ref{sec:data_selection}) and for
some of the spectral modelling (Section \ref{sec:diskbb}) we use a
multicolour disc blackbody, {\sc diskkbb} (Mitsuda et al. 1984). We
also use more advanced model taking into account the spectral
distortions from Comptonization in the disc and relativistic
corrections in Section \ref{sec:bhspec}. The disc photons are
Comptonized in optically thin plasma, for which we use thermal
Comptonization model {\sc thcomp} (Zdziarski, Johnson \& Magdziarz
1996; {\.Z}ycki, Done \& Smith 1999). We also include a Gaussian (with
energy constrained to be between 6--7~keV and $\sigma$ fixed at 0.5~keV)
and smeared edge (Ebisawa et al. 1991) to approximately describe the
reflected emission. The Galactic absorption is modelled as in Done et
al. (2004), i.e. with variable abundance cold gas {\sc varabs} with
abundances from Anders \& Ebihara (1982) and fixed column of
$N_H=4.7\times 10^{22}$ cm$^{-2}$ in all elements apart from Si and Fe,
which had equivalent H column (assuming ISM abundances) of $16.4\times
10^{22}$ and $10.9\times 10^{22}$ cm$^{-2}$, respectively, as
determined from the {\it Chandra} Transmission Grating results of Lee
et al. (2002).

We assume the black hole mass of 14 M$_{\odot}$ (Harlaftis \& Greiner
2004), the disc inclination of 66$^\circ$ (Fender et al. 1999) and the
distance to the source of 12.5 kpc (Rodr{\'{\i}}guez \& Mirabel 1999),
unless stated otherwise.

\section{Selecting Disc--Dominated spectra}
\label{sec:data_selection}

The derived disc temperature and luminosity are strongly
model-dependent where the X-ray tail constitutes more than $\sim$20
per cent of the total bolometric power (e.g. Kubota \& Done 2004).
Hence it is very important to select only disc-dominated spectra to
determine the black hole spin. We aim to chose only the spectra with
less then 15 per cent contribution from Comptonization to the
bolometric flux.

Another issue with data selection is the rapid variability seen in
GRS~1915+105, where the spectrum switches rapidly between states
with strong disc-like spectra (states A and B) and states with
strong high-energy emission (state C; see Belloni et al. 1997,
2000). Spectra accumulated over long timescales (e.g. a single {\it
RXTE} pointing, which is typically a few thousand seconds long) are
quite often a mixture of states, so cannot be averaged together.
Therefore, we can either extract spectra on much shorter timescales,
shorter than the spectral variability timescale, or try to find
longer periods with little variability.

\subsection{Long timescale spectra}
\label{sec:long_timescale_spectra}

We first investigated longer timescales by accumulating spectra over
entire pointed observations (designated by an identifier, obsid). We
selected only stable data intervals with less then 5 per cent rms
variability in light curves with 16-s resolution. There are some
observations in which the source was only in the disc-like state A
or B with very little variability. These were classified as
variability states $\phi$ and $\delta$ by Belloni et al. (2000).
However, they do not show necessarily show disc-dominated spectra.
We use the spectrum from obsid 10408-01-20-00 as an example of
variability class $\phi$ to illustrate the difficulties in fitting
these spectra.

\begin{figure}
\begin{center}
\begin{tabular}{c}
\leavevmode \epsfxsize=4.5cm \epsfbox{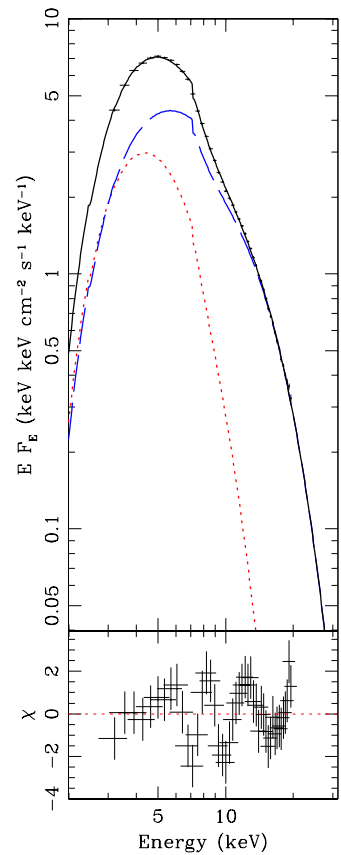}
\end{tabular}
\end{center}
\caption{The apparently disc dominated spectrum of variability class
$\phi$ (obsid 10408-01-20-00). The best fitting spectral
decomposition (using {\sc diskbb} and {\sc thcomp}) instead gives a
low-temperature Comptonization component (long-dashed curve) as well
as a disc (dotted curve). The total spectrum (solid curve) also
includes a Gaussian line (short-dashed) and iron edge feature to
roughly model the reflected emission. The lower panel shows
residuals, which are dominated by traces of highly ionized
absorption (Kotani et al 2000; Lee et al 2001).}
\label{fig:lt20}
\end{figure}

We extracted these data (all PCUs, all layers), together with
background and response using {\sc ftools} 5.3 using standard
extraction criteria.  We added a 1 per cent systematic error in each
energy channel to account for the residual uncertainties in the
response. We fitted them with the model described in Section
\ref{sec:model}, using {\sc diskbb} for the disc emission. Fig.
\ref{fig:lt20} shows our best-fitting deconvolution of this
spectrum, together with residuals which are dominated by a known
resonance absorption line (Kotani et al. 2000).

The continuum curvature is best described by a weak, low-temperature
disc plus substantial amounts of Comptonized emission than by a disc
alone. This could simply represent the effects of distortion by
Compton scattering within the disc itself, but such strong
distortions are not predicted by the best current disc models. These
remain rather close to a {\sc diskbb} shape as long as the
dissipation of gravitational potential energy occurs at more than a
few optical depths within the disc (Davis et al. 2005).

Such low-temperature Comptonization contrasts with the usual
spectral decomposition for the high/soft state in black hole
binaries, where the spectrum can be dominated by the disc, with a
weak Comptonized emission with photon spectral index of $\Gamma\sim
2$. However, high luminosity black hole binaries can also show an
alternative type of spectrum, where the spectrum below 20~keV can be
characterized by Comptonization with a temperature of 10--30~keV,
together with the disc emission (very high state/steep power law
state e.g. Kubota et al 2001; Kubota \& Done 2004).  The temperature
inferred here is much lower, at $kT_e\sim 3$~keV, but it could be a
more extreme version of this state.

The energy dependence of the variability gives another argument for
the reality of this low-temperature Comptonization component.
Zdziarski et al. (2001; 2005) found similar spectral decompositions
(disc plus low-temperature Comptonization) for apparently disc-like
spectra from the $\omega$ state of GRS 1915+105. They show that the
rms variability spectrum for these data increases with energy, as
predicted from models of a relatively stable disc and a more
variable Comptonization component. Thus, it seems likely that even
these apparently disc-like spectra are distorted by a (low
temperature) Comptonization component. Since the disc luminosity and
temperature can be significantly distorted in spectra with strong
Comptonization (Kubota \& Done 2004; Done \& Kubota 2006), then
these data cannot be used to give robust estimates of the disc
properties.

We re--examined all the spectral fits of Done et al. (2004), which
used all available {\it RXTE} observations before the end of Epoch 4
(up to 2000 May 11) and none of these had Comptonization which
contributed less than 15 per cent of the bolometric flux. This was
also the case with these data split into 128-s spectra. None of these
data fit the criteria for being disc dominated, so none of them are
suitable for disc spectral fitting.

\subsection{Short timescale spectral binning}

In Section \ref{sec:long_timescale_spectra} we have shown that there
are no disc-dominated spectra (with Comptonization contribution less
then 15 per cent) in the (stable) data accumulated on timescales longer
than 128 s. The apparently disc-shaped spectra are instead dominated by
low-temperature Comptonization. Therefore, we look at shorter
timescales of 16 s (the timing resolution of Standard-2 data) and check
if we can find disc-dominated spectra in the periods of larger
variability on longer timescales, which were excluded from our previous
analysis. As there still can be substantial variability over timescales
shorter than 16 s (Belloni et al. 2000), we select only those
variability classes where the characteristic switches between spectral
states are rather slow (classes $\beta$ and $\kappa$ of Belloni et al.
2000). We selected 16 pointed observations of these, as detailed in
Table \ref{tab:obsids}.

\begin{table*}
\begin{tabular}{cl}
\hline
Class & Observation ID \\
\hline
$\beta$ & 10408-01-10-00, 10408-01-21-00, 20402-01-43-00,
20402-01-43-02, 20402-01-44-00, 20402-01-45-00 \\
& 20402-01-45-03, 20402-01-46-00, 20404-01-52-01, 20402-01-53-00, 20402-01-59-00\\
$\lambda$ & 10408-01-37-00, 10408-01-38-00, 20402-01-36-00,
20402-01-36-01, 20402-01-37-01 \\
\hline
\end{tabular}
\caption{Observation IDs (obsids) from variability classes $\beta$
and $\kappa$ (Belloni et al. 2000) where transitions between
apparently disc dominated and Comptonized spectra are slow enough to
be potentially resolved by the 16-s time resolution data from
Standard-2 data.} \label{tab:obsids}
\end{table*}

\begin{table}
\begin{tabular}{cccc}
\hline
Observation ID & Start time (s) & Stop time (s) & Selection\\
\hline
10408-01-10-00     & 75749539  & 75749555\\
(Class $\beta$)    & 75755491  & 75755507\\
                   & 75755507  & 75755523\\
                   & 75756499  & 75756515\\
                   & 75756947  & 75756963 & (b)\\
\\
10408-01-21-01     & 79325363  & 79325379\\
(Class $\beta$)    & 79325395  & 79325411\\
                   & 79325459  & 79325475\\
                   & 79325475  & 79325491\\
                   & 79325491  & 79325507\\
\\
10408-01-38-00     & 87288867  & 87288883\\
(Class $\lambda$)  & 87289779  & 87289795\\
                   & 87290083  & 87290099\\
                   & 87290131  & 87290147\\
                   & 87293843  & 87293859\\
                   & 87293891  & 87293907\\
                   & 87293923  & 87293939\\
                   & 87293971  & 87293987\\
                   & 87295075  & 87295091\\
                   & 87295987  & 87295991 & (c)\\
                   & 87296051  & 87296067\\
                   & 87296179  & 87296195\\
                   & 87296211  & 87296227\\
                   & 87299347  & 87299363\\
                   & 87299363  & 87299379\\
                   & 87300275  & 87300291\\
                   & 87300387  & 87300403\\
\\
20402-01-45-03     & 116410403 & 116410419\\
(Class $\beta$)    & 116410419 & 116410435\\
                   & 116410435 & 116410451\\
                   & 116417059 & 116417075 & (a)\\
\\
20402-01-59-00     & 124956003 & 124956019\\
(Class $\beta$)    & 124957043 & 124957059\\
\hline
\end{tabular}
\caption{Log of 16-s disc-dominated spectra with less than 5 per
cent variability, used for spectral fitting in Section
\ref{sec:spectra}. Start and stop times are in mission elapsed time,
which is number of seconds since 1994 January 1, 0$^{\rm h}$0$^{\rm
m}$0$^{\rm s}$ UTC.  They are represented by black points in Fig.
\ref{fig:colcol}(b). The last column indicates the three spectra
selected for spectral analysis in Section \ref{sec:spectra}, and
indicated by diagonal crosses in Fig. \ref{fig:lt}}
\label{tab:16s-log}
\end{table}

We extracted PCA spectra in 16-s segments from Standard-2 data for
each of the these observations, using {\sc ftools} 5.3. We used all
available PCUs and all the PCA layers and corrected the spectra for
background and dead-time effects. No systematic uncertainties were
added to the data as the short integration time gave statistical
uncertainties which were always larger than $\sim$1 per cent in each
channel. In order to ascertain the variability within 16-s
intervals, we also extracted corresponding Standard-1 light curves
with 0.125-s resolution, but no spectral resolution. We fitted each
spectrum in {\sc xspec} using the model described in Section
\ref{sec:model}, with {\sc diskbb} disc emission, and seed photon
temperature independent of the disc temperature. We obtained
acceptable fits, with reduced $\chi^2_\nu < 1.3$, for all $\sim$4000
spectra.

\begin{figure*}
\begin{center}
\begin{tabular}{cc}
\leavevmode \epsfxsize=7cm \epsfbox{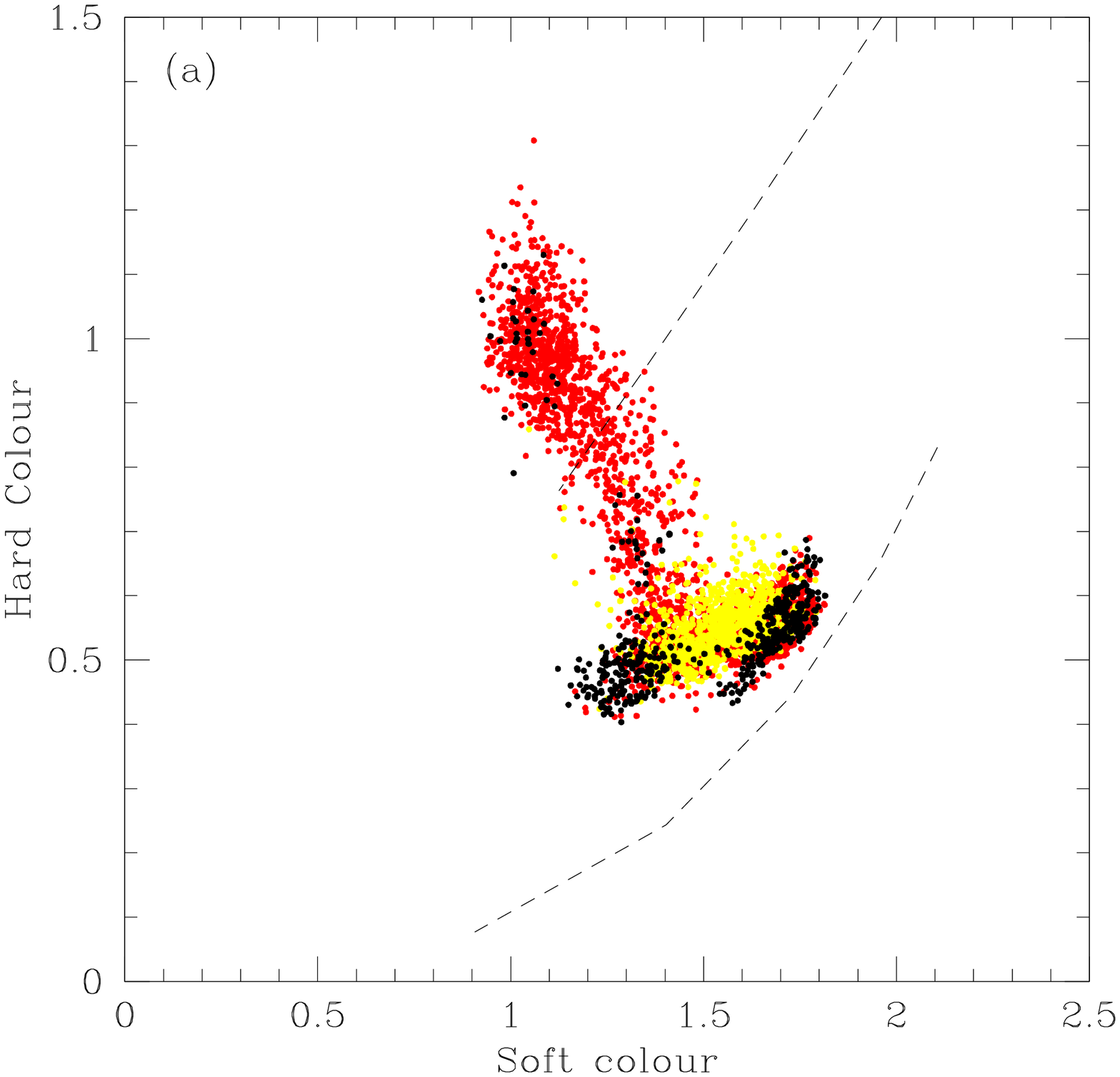} &
\epsfxsize=7cm \epsfbox{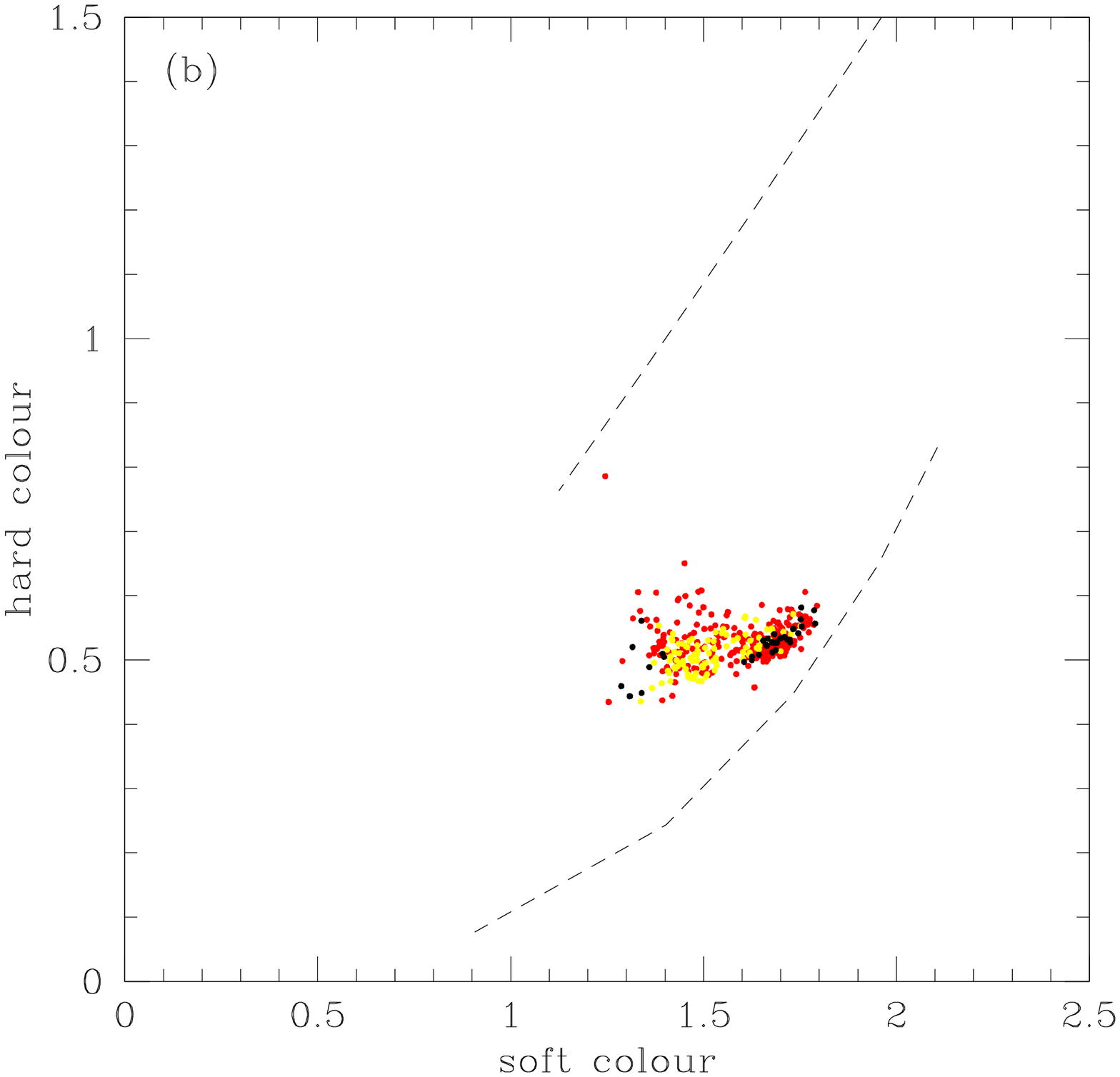}\\
\end{tabular}
\end{center}
\caption{Colour-colour diagrams for the spectra extracted in 16-s
intervals. Red points are spectra with 50 per cent or more rms
variability, yellow between 50 and 20 per cent rms, while black is less
than 5 per cent. (a) Colours calculated from all observations listed in
Table \ref{tab:obsids}. (b) A subset of the entire sample, limited to
the disc-dominated spectra, with less than 15 per cent of the
bolometric luminosity in the hard tail. The diagonal dashed line
indicates the range of colours from a power-law spectrum with index
varying from 1.5 (top right) to 3.0 (bottom left) while the lower curve
indicates the colours expected from a disc dominated spectrum at
different temperatures, derived from {\sc diskbb} with temperature
1--3~keV. We chose only disc-dominated spectra with less then 5 per
cent rms variability--black points in panel (b)--for detailed
analysis.} \label{fig:colcol}
\end{figure*}

Removing absorption from each model fit gives the best estimate of the
intrinsic spectrum of the source using these components.  This model
spectrum is then used to calculate intrinsic soft and hard colours
(i.e. corrected for the effects of absorption and the instrument
response), as in Done \& Gierli{\'n}ski (2003). These are formed from
the (energy) flux ratios of 4--6.4/3--4~keV and 9.7--16/6.4--9.7~keV,
for the soft and hard colours, respectively. We also use the unabsorbed
model spectrum to estimate the total bolometric luminosity from
0.01--1000~keV flux, for the distance of 12.5~kpc.

We use these fluxes to select data where the disc contributes more
than 85 per cent of the total bolometric luminosity.  However, not all
of these spectra are necessarily appropriate to use for the disc
analysis as the variability in GRS 1915+105 is extreme.  The source
can change significantly (in both spectral shape and normalization) on
timescales shorter than 16 s. The spectra accumulated over such
intervals can be severely distorted, even if they consist only of a
pure disc varying in temperature. Hence, we fold in an additional
selection criteria, which is that the rms variability during each 16-s
interval should be less than 5 per cent.  We use the 0.125-s
resolution Standard-1 light curves corresponding to each 16-s spectrum
to calculate the rms amplitude.

Fig.~\ref{fig:colcol} illustrates our data selection on a
colour-colour diagram. The dashed lines show the intrinsic colours
expected for a power law and pure disc spectrum (upper and lower
curves, respectively). Fig.~\ref{fig:colcol}(a) shows all the data,
where red, yellow and black points indicate an
rms variability of 50--20, 20--5 and less than 5 per cent,
respectively. Fig.~\ref{fig:colcol}(b) shows the same diagram for
those spectra where the disc is dominant (where the tail is less
than 15 per cent of the total bolometric luminosity). As expected,
these all lie fairly close to the spectrum of a disc (lower dashed
curve) as opposed to a power law (upper dashed straight line), but
some are still quite variable, so could be a mix of states
and/or disc temperatures. Hence we only use spectra where there is less
than 5 per cent rms variability (black points), leaving a total of 34
disc-dominated, steady spectra across 6 obsids with which we carry out
our analysis (Table \ref{tab:16s-log}).

\section{Disc dominated, steady spectra}
\label{sec:spectra}

\begin{table*}
\centering
\begin{tabular}{l|l|l|l|l|l}

\hline
Model component & Parameter & spectrum (a) & spectrum (b) & spectrum (c) \\
\hline {\sc diskbb} &  $T_{\rm in}$ (keV) & $1.337_{-0.002}^{+0.004}$ & $1.71_{-0.05}^{+0.02}$ & $1.92_{-0.05}^{+0.04}$\\
             & $R_{\rm in}$ (km) & $38.0_{-0.7}^{+1.3}$ &  tied & tied\\
{\sc thcomp} & $\Gamma$ & $2.53_{-0.22}^{+0.39}$ & $1.04_{-0.03*}^{+0.32}$& $1.25_{-0.24*}^{+0.29}$ \\
             & $kT_{e}$ (keV) & $3.75_{-0.55}^{+0.68}$ & $2.90_{-0.18}^{+0.13}$ & $2.59_{-0.10}^{+0.13}$ \\

             & $\chi^2_\nu$ & 133.6/113 \\
\hline {\sc bhspec} & $L_{\rm disc}/L_{\rm Edd}$ & $0.524_{-0.005}^{+0.007}$ & $0.96_{-0.15}^{+0.01}$ & $1.52_{-0.08}^{+0.22}$ \\
             & $a_*$  & $0.720_{-0.017}^{+0.009}$ & tied & tied \\
{\sc thcomp} & $\Gamma$ & $1.01_{-0*}^{+0.25}$ & $1.22_{-0.21*}^{+0.42}$ & $3.1_{-0.9}^{+0.7}$\\
             & $kT_{e}$ (keV) & $2.51_{-0.05}^{+0.17}$ & $3.28_{-0.23}^{+0.22}$  & $8.8_{-4.5}^{+91*}$ \\
             & $kT_{\rm bb}$ (keV) & (1.34) & (1.71) & (1.92) \\

             & $\chi^2_\nu$ & 143.8/113 \\
\hline
\end{tabular}
\caption{Best-fitting parameters for the 3 representative spectra
fitted simultaneously with the model consisting of the disc emission
and thermal Comptonization. The black hole mass of 14 M$_\odot$, the
disc inclination of 66$^\circ$ and the distance of 12.5 kpc were
assumed (where applicable). In the model with {\sc diskbb} the inner
disc radius, $R_{\rm in}$, is calculated from the {\sc diskbb}
normalization, without any colour or relativistic corrections. In
the model {\sc bhspec} the seed photon temperature was fixed at the
best-fitting values of the {\sc diskbb} model. The inner disc radius
in {\sc diskbb} and the spin parameter in {\sc bhspec} were tied
across the three spectra. An asterisk denotes the parameter reaching
its hard limit within the error bar. The $\chi^2_\nu$ quoted is for
all three spectra fitted together.} \label{tab:fits}
\end{table*}

\subsection{{\sc diskbb} model results}
\label{sec:diskbb}

Each disc dominated spectrum was refit using the same model as before
except that the seed photons for the Comptonization are now tied to
those of the disc since these spectra are disc dominated. The derived
disc parameters (temperature and flux) are used to plot a
luminosity-temperature ($L$-$T$) diagram as in GD04. We corrected the
{\sc diskbb} temperature and flux from each spectrum for the effects of
disc geometry and relativistic smearing (assuming a given spin and
inclination of 66$^\circ$), plus an additional correction to
incorporate a stress-free inner boundary condition (which is not
included in {\sc diskbb}). Then, we converted this to a luminosity
using the assumed distance of 12.5~kpc.  These corrected data points
can then be compared with the theoretically expected $L$-$T$ plot for
the assumed mass (14 M$_\odot$) and spin (correction factor
interpolated for an inclination of 66$^\circ$ from the tabulated values
of Zhang et al. 1997) of the black hole, for an assumed colour
temperature correction of 1.8 (Shimura \& Takahara 1995) to relate the
observed temperature which is distorted by Compton scattering to the
equivalent blackbody temperature.

\begin{figure*}
\begin{center}
\begin{tabular}{cc}
\leavevmode \epsfxsize=6cm
\epsfbox{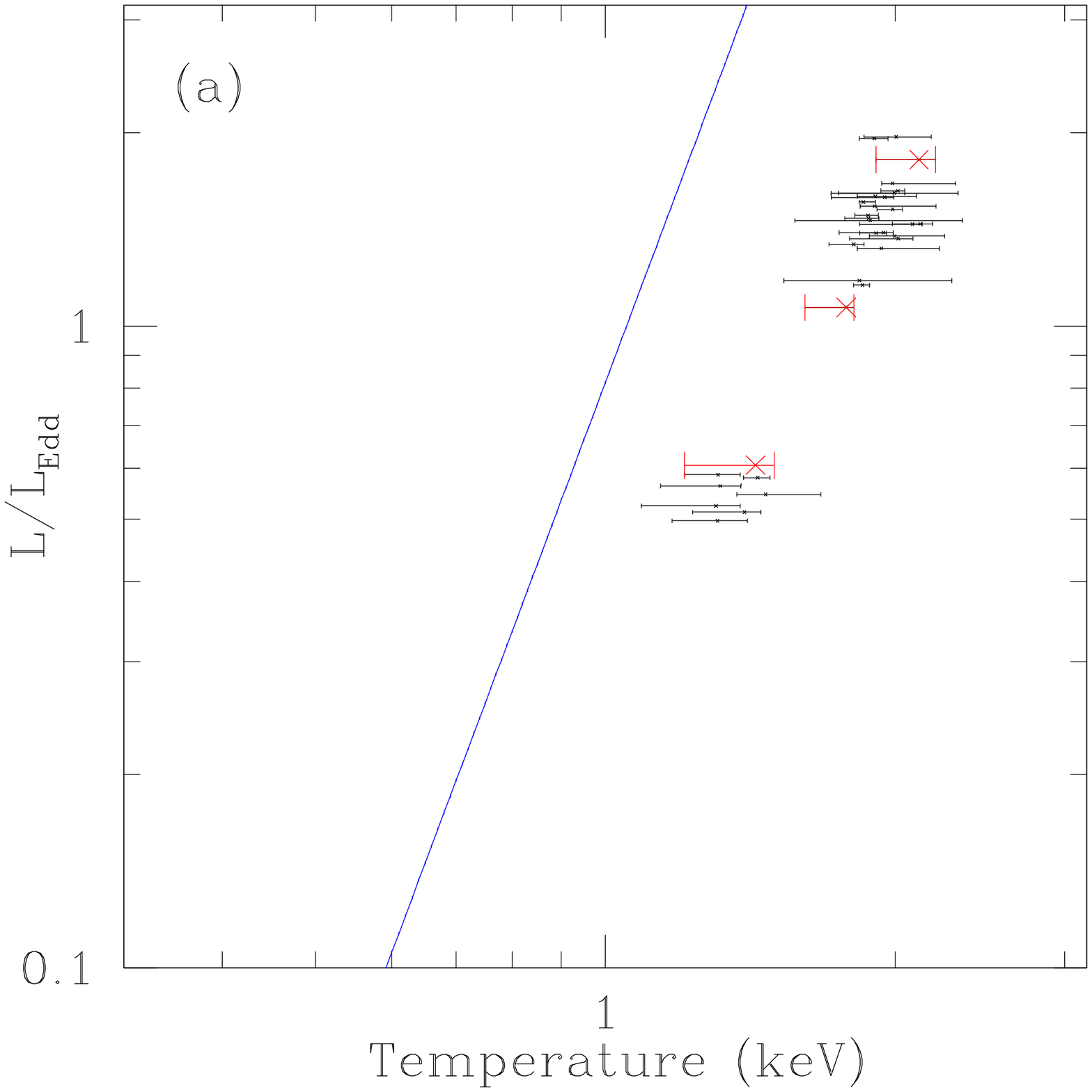} & \epsfxsize=6cm \epsfbox{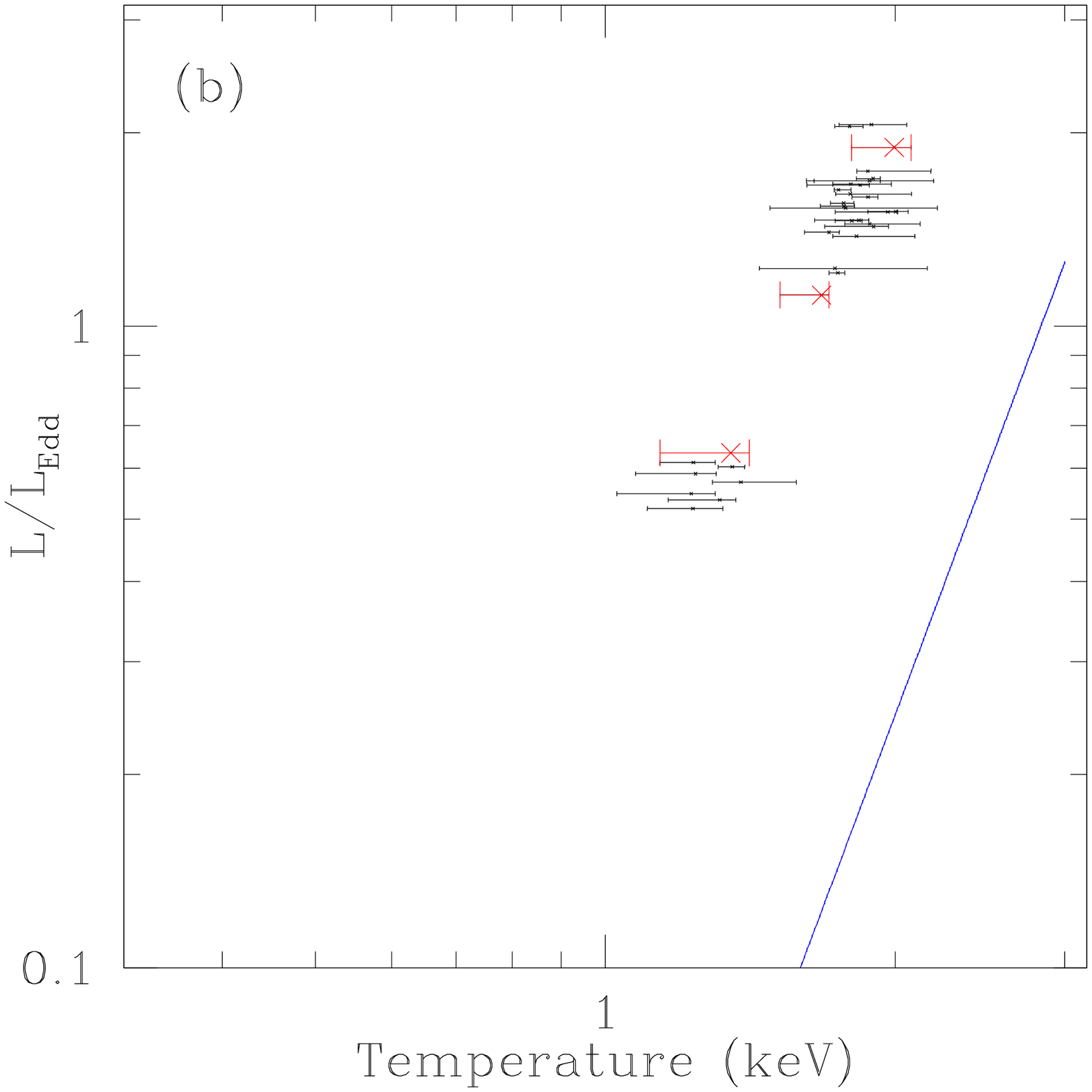}\\
\end{tabular}
\end{center}
\caption{Luminosity-temperature plots for our sample of disc-dominated
data with less than 5 per cent rms variability. (a) The intrinsic disc
luminosity and temperature (data points) derived assuming a
Schwarzschild black hole (Zhang, Cui \& Chen 1997) The solid line
represents the expected $L\propto T^{4}$ relation. (b) The same data
points, but corrected for maximal Kerr spin with the corresponding
$L\propto T^{4}$ relation expected for the much smaller disc inner
radius. Plainly the data lie between these two extreme values. The
points highlighted by diagonal crosses indicate the data selected for
further spectral fitting. These span the entire luminosity range and
are representative of the source behaviour so can be used to determine
the black hole spin.} \label{fig:lt}
\end{figure*}

Figs~\ref{fig:lt}(a) and (b) show the corrected data and expected model
$L$-$T$ relation for zero and maximal spin. Plainly the data follow an
approximate $L\propto T^4$ relation as expected from constant radius
disc, and equally plainly the data suggest the spin is intermediate,
rather than zero or maximal.

We can explicitly test how well the source follows the $L\propto T^4$
relation i.e. whether the emission is consistent with a constant inner
radius disc (with constant colour temperature correction) by fitting
several of the disc spectra simultaneously within {\sc xspec}, with the
disc inner radius tied across all data sets. We select 3 of the
disc-dominated spectra, shown by the diagonal crosses in
Fig.~\ref{fig:lt}, encompassing the spread in luminosity seen from the
disc. When fit separately these give a total $\chi^2_\nu=127.2/111$,
while when fit simultaneously, with the {\sc diskbb} normalization
($\propto r^{2}$) tied across all 3 data sets gives 133.6/113. Thus the
3 data sets are completely statistically consistent with a constant
radius, constant colour temperature corrected disc spectrum. The
results are shown in Table \ref{tab:fits}.

\subsection{{\sc bhspec} model results}
\label{sec:bhspec}

\begin{figure*}
\begin{center}
\leavevmode \epsfxsize=10cm \epsfbox{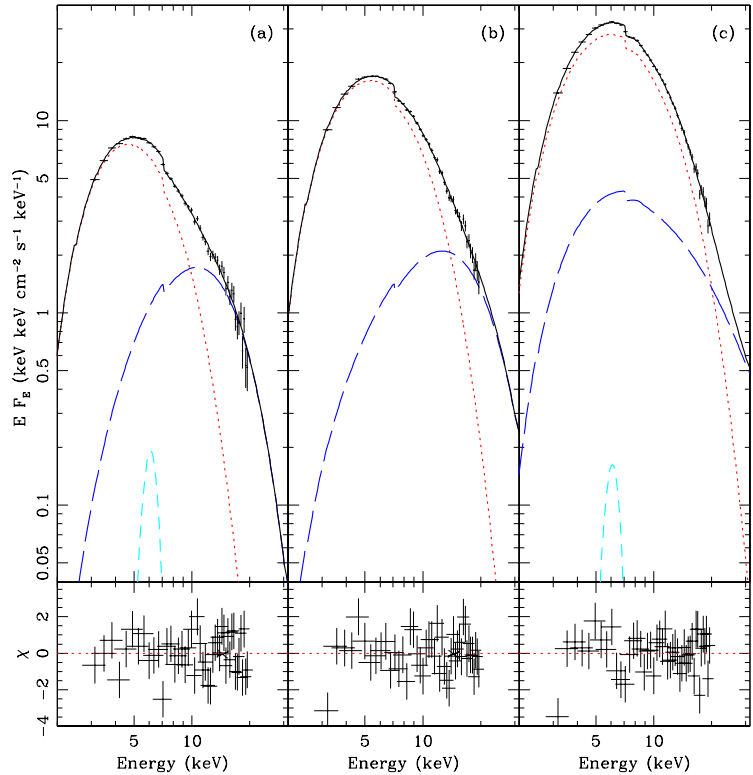}
\end{center}
\caption{Individual fits of {\sc bhspec} of the 3 ultrasoft spectra
which encompass the observed range in luminosity of disc dominated
spectra. The 3 plots beneath each spectrum show the respective
residuals for each fit. The data (black points) are fit by a model
consisting of the {\sc bhspec} disc spectrum (red line), its
Comptonized emission (blue line) and Gaussian iron line (cyan line)
giving a total spectrum (black line).} \label{fig:specbh}
\end{figure*}

The spectral model {\sc diskbb} assumes that each point of the disc
emits as a blackbody. In practice, the local spectrum is more complex.
At low frequencies true absorption opacity dominates and the emission
is blackbody, but for higher frequencies, electron scattering opacity
is larger leading to departures from local thermodynamic equilibrium
and producing a modified blackbody spectrum (see e.g. Shakura \&
Sunyaev 1973).  Also, the absorption opacity has significant
contributions from free-bound (photo-electric edges) as well as
free-free which again give frequency dependent effects on the radiative
transfer. {\sc bhspec} (now publicly available\footnote{{\sc bhspec}
can be downloaded from
http://heasarc.gsfc.nasa.gov/docs/xanadu/xspec/newmodels.html} as an
{\sc xspec} table model) utilizes a self-consistent calculation of the
spectrum including all these radiative transfer effects as a function
of inclination (Davis et al. 2005). {\sc bhspec} also incorporates the
relativistic effects and inner boundary condition directly in the
model.

We use the same 3 spectra as before and fit them using {\sc bhspec}
as the disc component (Fig.~\ref{fig:specbh}), assuming the
viscosity parameter $\alpha=0.01$ (Davis et al. 2006). We fixed the
seed photon temperature in the Comptonization model to that of the
previous {\sc diskbb} fits, as {\sc bhspec} is parameterized by mass
accretion rate rather then by temperature. This gives
$\chi^2_\nu=143.8/113$ and $a_*$ = 0.720$_{-0.017}^{+0.009}$. The
fit results are shown in Table \ref{tab:fits}.

We note that the fit using {\sc bhspec} is marginally worse than
the fit using {\sc diskbb}. This is because the relativistic
corrections included in {\sc bhspec} give a slightly different
spectral shape than {\sc diskbb} and the data slightly
prefer a Wien shape to the relativistically distorted Wien.  However,
at this level there are other approximations which can also affect the
spectrum, such as the amount of returning radiation illuminating the
disc and the spectral shape of the seed photons assumed for the
Comptonized emission (see also Davis et al. 2006).

\subsection{System Parameter Uncertainties}

The fits above all assume a distance of 12.5~kpc. However, this is
not well determined, and estimates range from this to as low as
7~kpc, though 11~kpc is the only distance which is compatible with
all the constraints (see e.g. Zdziarski et al. 2005).  We
investigate the effect this uncertainty has on the spin estimates by
refitting the spectra with these distances within the {\sc bhspec}
model.  We find that the distance simply trades off against the
inferred spin and luminosity giving $a_*$ = 0.990$_{-0.005}$ (there
is no upper limit as the {\sc bhspec} model table we use only
includes values of spin up to 0.99) and 0.819$_{-0.007}^{+0.005}$,
for the distance of 7 and 11 kpc, respectively.

The most likely mass of the source is 14$\pm$4 M$_{\odot}$ (Harlaftis
\& Greiner 2004). It is highly unlikely that this mass is an
overestimate as the source would be even more super Eddington at lower
masses, however an underestimate is possible if the jet axis is
misaligned.  To address this uncertainty we tested the effect on the
spin parameter with mass frozen at the upper limit of 19M$_{\odot}$.
Fitting the same spectra as before a spin value of $0.88\pm 0.01$ is
obtained.

%renorm_new_bestfit/_2
The inclination of the jet has remained constant to within a few
degrees for several years (Fender et al. 1999). This motivates the
belief that the jet is perpendicular to the axis of the disc. The value
of Fender et al. (1999) of (66$\pm$2)$^{\circ}$, as determined through
MERLIN observations, provides the best estimate for the inclination of
the system.  However, the upper limit of the eclipsing angle,
determined to be $79^{\circ}$ (Greiner et al. 2001) yields a useful
lower limit to the spin through fitting the same three spectra of
$0.10^{+0.04}_{-0.07}$.

\section{Alternative spin determinations}

\begin{figure}
\begin{center}
\leavevmode \epsfxsize=7cm \epsfbox{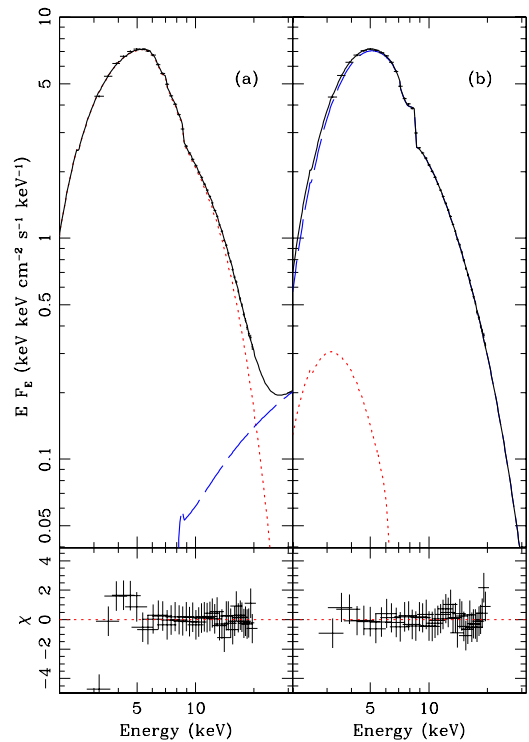}
\end{center}
\caption{Fits for obsid 10408-01-20-00 using {\sc bhspec} and {\sc
comptt} with the electron temperature fixed at 50 keV (a) and free
(b). The plots beneath each spectrum show the respective residuals
for each fit. The data (black points) are fit by a model consisting
of the {\sc bhspec} disc spectrum (dotted curve), its Comptonized
emission (dashed curve) giving a total spectrum (solid line) without
an Gaussian iron line in either case.} \label{fig:mc_comp}
\end{figure}

\subsection{Disc spectral fitting}

McClintock et al. (2006) similarly study the spin of GRS~1915+105
from a much larger sample of {\it RXTE} data, but find rather
different results. Their inferred spin is luminosity-dependent, with
low-luminosity data predicting maximal spin, while at higher
luminosities, $\ga$ 0.4 $L_{\rm Edd}$, the measured spin gradually
decreases, with their high-luminosity spin measurements being
comparable with our results. They argue that the low luminosity
results are more robust due to theoretical uncertainties on the disc
structure at high luminosities. It is certainly true that there are
many uncertainties on calculating disc structure at luminosities
close to Eddington and above.  However, the lower luminosity spectra
are affected by uncertainties in spectral modelling. McClintock et
al. (2006) used several different descriptions of the Comptonized
emission (power law, {\sc comptt} and exponentially cutoff power
law), but none of these allowed for the possibility of a temperature
as low at $\sim 3$~keV (see Section
\ref{sec:long_timescale_spectra})

We illustrate the limitations of this approach by refitting one of
their spectra (obsid 10408-01-20-00, as in Section
\ref{sec:long_timescale_spectra}) using a model very similar to
theirs, consisting of {\sc bhspec} and {\sc comptt}, Galactic
absorption, broad absorption line, smeared edge, and an additional
narrow edge at $\sim$8 keV. Following McClintock et al. (2006) we
have fixed the seed photon temperature for Comptonization at 2.0 keV
and electron temperature at 50 keV. We obtained a good fit
($\chi^2_\nu$ = 38.0/36), where the spectral decomposition is
dominated by the disc (Fig. \ref{fig:mc_comp}a). The high disc
temperature then requires a high spin. However, a better fit is
obtained by allowing the electron temperature to be free, giving
$\chi^2_\nu=13.0/35$ (showing the effect of the systematic rather
than statistical uncertainties). The spectrum is then dominated by
the low-temperature Comptonized component (similar to the one shown
in Fig. \ref{fig:lt20}), where the disc contributes only 10 per cent
of the flux and is at much lower temperature (Fig.
\ref{fig:mc_comp}b). The derived Comptonization temperature is
$3.25^{+0.41}_{-0.58}$~keV (where the limits are 90 per cent
confidence on one free parameter, i.e. $\Delta\chi^2/\chi^2_\nu$ =
2.7, or $\Delta\chi^2$ = 7.27). The upper limit is far below the
values assumed by McClintock et al. (2006). The spin is not
constrained now (see Table \ref{tab:mcfits}) and thus is consistent
with our value derived in Sec. \ref{sec:spectra}. This shows that
the derived spin is dependent on the details of the data analysis,
contrary to the robustness claimed by McClintock et al. (2006).

\subsection{Broad iron line}

A completely different method to potentially determine the spin is via
relativistic broadening of the iron line and Compton reflection features
(see e.g.  the review by Fabian et al. 2000). Unlike the disc spectrum
method, this depends on having a strong X-ray tail in order to pick out
its reflection.  Results from this are currently confused, mostly
because the line is very variable. Most observations show that the line
is not highly smeared, so the inner radius of the line emitting material
is $\gg R_g$ and hence cannot constrain the spin (e.g. Martocchia et al.
2002, 2004; Sobolewska \& $\dot{\rm Z}$ycki 2003). However, there is one
{\it BeppoSAX} observation where the derived line is extremely smeared,
with the inner disc radius $R_{\rm in}=1.4^{+3.7}_{-0.2}R_g$, so
requires a highly spinning black hole (Martocchia et al. 2004). This
changing smearing on the line could be due to real changes in the disc
inner radius, as required in the limit cycle instability models for this
source where the disc empties and refills (e.g. Belloni et al.
1997). It could also be due to changes in ionization of a constant
radius disc. A third possibility is that the extreme smearing inferred
form some spectra are actually artifacts of ionized absorption (Done \&
Gierli\'{n}ski 2005) which is known to be present in this source (Kotani
et al. 2000; Lee et al. 2002).  None of these are mutually exclusive,
but whatever the cause, our best estimate of the spin of $a_*=0.7$
implies $R_{\rm in}=3.4R_g$ so it is consistent (within the
uncertainties) on even on the most extreme inferred smearing of the iron
line.

\subsection{Quasi-Periodic Oscillations}

Yet another constraint on the properties of the innermost parts of the
accretion flow is via the variability behaviour on short timescales.
There are a number of quasi periodic oscillations (QPOs) which can be
seen in the power spectra of the Galactic black hole binaries. The
strongest of these is generally the low-frequency (LF) QPO, which moves
in frequency, typically between 0.5--10~Hz. The properties of this
feature are correlated with the spectral shape.  In general it is seen
when the spectrum consists of both a strong disc and strong X-ray tail,
and the LF QPO frequency increases as the spectrum softens (e.g. Muno,
Morgan \& Remillard 1999). This feature can be both strong and sharp,
so giving a clear observational diagnostic. The problem is that there
is no widely accepted theory for QPO generation, though all current
models require some moving characteristic radius in the disc to
generate the changing frequency of the LF QPO. One of the promising
approaches is to identify the LF QPO frequency with the precession
frequency of a vertical perturbation in the flow, which depends on both
the characteristic radius at which the QPO is produced, and black hole
spin (Stella \& Vietri 1999).  Muno et al. (1999) show that the LF QPO
spans 0.5--10~Hz in GRS 1915+105, so the maximum frequency must be
produced by a disc with radius larger or equal to the last stable
orbit. This only limits the spin to be $a_* > 0.3$ (using the full
expression for Lens-Thirring precession in Merloni et al. 1999) or $a_*
> 0.2$ if the LF QPO is at twice the precession frequency.

\begin{table*}
\centering
\begin{tabular}{l|l|l|l}

\hline
Model component & Parameter & (a) & (b)                \\
\hline
{\sc bhspec} &  $L_{\rm disc}/L_{\rm Edd}$  & $0.344_{-0.004}^{+0.003}$ & $0.03_{-0.02}^{+0.03}$\\
& $a_{*}$ & $0.990^{+0*}_{-0.002}$ & $0.91_{-0.91*}^{+0.08*}$\\
{\sc comptt} &$T_{0}$ (keV)& (2.0) & (1.0) \\
& $kT_e$ (keV) & (50) & $3.25_{-0.48}^{+0.41}$\\
& $\tau$ & $1.6_{-1.2}^{+5.6}$ & $2.6_{-0.2}^{+0.8}$\\
             & Normalization  & $8.3_{-0.9}^{+15}\times10^{-4}$ & $1.33_{-0.15}^{+0.25}$\\

             & $\chi^2_\nu$ & 38.0/36 & 13.0/35  \\

\hline
\end{tabular}
\caption{Best-fitting parameters with errors for the two fits of
obsid 10408-01-20-00 with (a) electron temperature fixed as in
McClintock et al. (2006) and (b) free to vary. Note that the seed
photon temperature ($T_{0}$) is fixed to the value of McClintock et
al. (2006) at 2 keV for (a) but we use 1 keV in our model (b) as
this is more appropriate for the much lower luminosity inferred for
the disc. Values in parentheses denote fixed parameters. An asterisk
denotes the parameter reaching its hard limit within the error bar.
As the reduced $\chi^2$ in model (b) was much less than one, the
errors were calculated for $\Delta\chi^2/\chi^2_\nu$ = 2.7, i.e.
$\Delta\chi^2$ = 7.27.} \label{tab:mcfits}
\end{table*}

A more direct measure of black hole spin may be given by the
high-frequency (HF) pair of QPO features which seem fairly fixed in
frequency and in a 3:2 harmonic relation (though in GRS 1915+105 the
frequency ratio are 3.36:2, so not exact; Strohmayer 2001).  Harmonics
are most naturally associated with a resonance in the disc, but again
there are several possibilities for this resonance so as yet no unique
way to translate these frequencies into black hole spin. T\"{o}r\"{o}k
et al. (2005) review the possible orbital resonances in a thin disc,
and highlight the parametric resonance between vertical and radial
epicyclic modes in strong gravity as these rather naturally give a 3:2
ratio (Klu{\'z}niak \& Abramowicz 2002). This identification implies
$a_*$ = 0.9--0.98 for the observed HF QPOs in GRS 1915+105. However,
there are other resonances even in a thin disc which can also give a
3:2 ratio, though these generally predict other harmonics as well.
Depending on which of these is chosen as the origin for the HF QPOs
gives a value for the spin spanning the whole range from $a_*$ =
0--0.998 (T\"{o}r\"{o}k et al.  2005). Another, perhaps more attractive
possibility, is that the HF QPO is a resonance in a {\em thick} disc,
where there are additional modes available from compression and
extension of the disc scale height (breathing modes). Blaes, Arras \&
Fragile (2006) show that a `natural' 3:2 resonance forms between the
vertical epicyclic mode and a `breathing' mode (Blaes, Arras \& Fragile
2006). This identification gives much weaker constraints on spin than
the parametric resonance.

The dispersion of values for the spin does not look promising and
obviously all the techniques covered here have differing strengths and
weaknesses.  For the spectral methods, the physical origin of the
features (disc or iron line) is clear, and the uncertainties are due to
the observations, or rather how to unambiguously isolate the disc
emission or relativistic line profile in a complex spectrum.
Conversely, the timing methods are unambiguous observationally, but
lack a theoretical model to uniquely associate the measured frequencies
with black hole spin. Nonetheless, from this brief review it is clear
that only potentially significant conflict with moderate ($a_* = 0.7$)
spin in GRS 1915+105 for a reasonable distance is the parametric
resonance interpretation of the HF QPO which requires $a_* \ga 0.9$,
while the newer interpretation of this as the resonance between the
vertical epicyclic and `breathing' modes is consistent with any spin.

We note that the parametric resonance between the radial and
vertical epicycles interpretation of the HF QPOs also gives a much
higher spin for GRO J1655--40 and XTE J1550--564 of $a_* \sim 0.96$
(T\"{o}r\"{o}k et al 2005) than are derived from their disc
spectrum, $a_* \sim 0.7$ (Shafee et al. 2006) and $a_* \sim 0.1$
(Davis et al. 2006), respectively. There are no issues in these
sources with the strong variability or interstellar absorption, and
the distances are rather better determined, so there seems to be a
clear discrepancy between the results of the parametric resonance
interpretation of the HF QPOs and the disc spectra. Again, this
conflict is removed if the HF QPO is instead a resonance between the
vertical epicyclic and breathing modes (Blaes et al. 2006).

\section{Black hole spin and jet power}

One of the big questions in understanding accretion flows is how the
jet is powered, whether it predominantly taps the gravitational
potential or whether it predominantly taps the spin energy plausibly
via the Blandford-Znajek mechanism. The possibility of spin powered
jets has given rise to persistent speculation in the literature that
relativistic jets require a maximally spinning black hole. GRS~1915+105
is the most powerful X-ray binary jet source in our Galaxy, so is a key
object for testing models of jet formation. Our fits to the
disc-dominated spectra from this source clearly show that the spin is
unlikely to be maximal, though it is substantial ($a_*$ = 0.7--0.8 for
any reasonable distance estimate). The same technique (simultaneous
fitting of a series of disc dominated spectra at differing
luminosities) gives non-maximal spins ($a_*$ = 0.1--0.8) for four other
Galactic black holes (Davis et al. 2006; Shafee et al. 2006). These
objects also show radio jets, sometimes with superluminal motion, so it
is clear that these results imply that powerful jets do not require
maximally spinning black holes.

The next question is whether spin is needed at all in producing a jet,
i.e. whether the jet can be purely powered by gravity. Observationally
it is clear that the jet power scales with accretion rate (Gallo,
Fender \& Pooley 2003). Since all the Galactic black holes are more or
less consistent with the same ratio of radio to X-ray power (which
traces the jet to accretion power) at a given $L/L_{\rm Edd}$ (in the
hard spectral state), then the observed range of $a_*$ = 0.1--0.8
implies that the ratio of jet to accretion power does not depend
strongly on spin. This immediately favours gravity powered jets, as
spin powered jet models generically depend rather strongly on spin!

The best current jet models are those which are produced in numerical
simulations of the jet/accretion flow. These include the
self-consistent, magnetically generated stresses and produce jets and
outflows without additional physics. These show in general that the jet
has two components, a matter dominated, funnel wall jet and an
electromagnetic Poynting flux jet (McKinney 2005; Hawley \& Krolik
2006). The electromagnetic jet is probably highly relativistic with
bulk Lorentz factor $\gg 1$, and is very strongly dependent on $a_*$,
indicating that this may be partly (or perhaps even mostly) powered by
the black hole spin (McKinney \& Gammie 2004).  By contrast, the funnel
wall jet is less relativistic (perhaps only Lorentz factors $\le
$~2--3) and is much less dependent on black hole spin (McKinney 2005;
Hawley \& Krolik 2006).  The potential power of the matter jet relative
to the accretion power increases by only a factor 2 between $a_*$ = 0.5
and 0.95 (Hawley \& Krolik 2006). Such a small change is unlikely to
result in much scatter in the observed ratio of radio to X-ray
luminosity, in contrast with the factor 11 for the Poynting flux jet
(Hawley \& Krolik 2006). While we caution that the simulations do not
currently include radiation, so cannot yet be unambiguously connected
to observations, it seems that the funnel wall, matter dominated jet,
powered predominantly by the gravity of the accretion flow, matches
rather well to the properties of the jets in Galactic black hole
binaries in that it can accommodate the observed constancy of radio to
X-ray flux at a given $L/L_{\rm Edd}$ from objects with the variety of
spins inferred here (see McKinney 2005).

The moderate spins derived from the disc spectra (as opposed to the
near maximal spins derived from the HF QPOs: T\"{o}r\"{o}k et al. 2005)
also match with the theoretical predictions for the birth spin
distribution of black holes, as the pre-supernovae core before stellar
collapse is slowly rotating, and spin up from captured fallback of
material is countered by angular momentum loss in gravitational waves
during the formation process (see e.g. the review by Gammie, Shapiro \&
McKinney 2004). Spin-up through accretion during the lifetime of the
binary is limited in most systems as the companion mass is generally
smaller than the black hole mass (King \& Kolb 1999).

\section{Conclusions}

We have searched the {\it RXTE} spectra from GRS 1915+105 to find
the rare disc-dominated states. These can only be found at the
shortest PCA (Standard-2 data) timescales of 16 s, as such spectra
are seen only during the unique limit cycle variability of this
source. We fit these with the best current models of the accretion
disc spectrum, which include full radiative transfer through a solar
abundance atmosphere as well as the full relativistic dissipation
(with stress free inner boundary condition) and ray tracing to
incorporate the relativistic effects on propagation of the emission
(Davis et al. 2005; Davis \& Hubeny 2006). This gives the spin of
the black hole as $a_* \sim 0.7$, assuming the best current estimate
for distance and binary system parameters.  This, together with
spins determined using this method for other galactic black holes
gives a distribution of $a_* \sim$ 0.1--0.8 i.e. low-to-moderate
spin. Moderate (as opposed to maximal) spins are consistent with
theoretical expectations of the spin distribution of galactic black
holes, and with the jet properties, but conflict with the
high-to-maximal spin inferred (often for the same objects) from the
broad iron line profile and parametric vertical-radial epicyclic
resonance interpretation of the HF QPOs. While the broad line
profile may be distorted by highly ionized absorption lines (Done \&
Gierli\'{n}ski 2006), the discrepancy between the spins derived from
disc spectra and those derived from the vertical-radial epicyclic
resonance interpretation of the HF QPO appears robust. Given the
apparent simplicity of the accretion disc behaviour, it seems most
likely that the HF QPO should instead be associated with another
resonance in the disc, plausibly the vertical epicycle-breathing
resonance identified by Blaes et al. (2006).

\section*{Acknowledgements}

We thank Jean-Pierre Lasota and Steve Balbus for useful discussions on
the difference between spin and gravity powered jets. We also thank
Julian Krolik and Jon McKinney for their help in understanding how to
interpret the results of the MRI simulations. The editors of MNRAS
probably deserve the undying gratitude of the readers for objecting to
the wretched pun of the original title of `The spin of GRS 1915+105:
why do we Kerr?'

\label{lastpage}

\end{document}